\documentclass[pra,aps,twocolumn,showpacs,10pt,floatfix]{revtex4-1}
\usepackage{amsmath,amssymb,graphicx,bm,color,ulem}
\usepackage{amsfonts,amsthm}

\begin{document}

\title{Comment on ``Spatiotemporal torquing of light"}

\author{Miguel A. Porras}
\address{Grupo de Sistemas Complejos, ETSIME, Universidad Politécnica de Madrid, Rios Rosas 21, 28003 Madrid, Spain}

\begin{abstract}
The recent paper Phys. Rev. X. 14, 011031 (https://doi.org/10.48550/arXiv.2307.01019) includes an appendix that casts doubts on the validity the theory of the transverse orbital angular momentum of spatiotemporal optical vortices (STOVs) in Prog. Electromagn. Res. 177, 95 (https://doi.org/10.48550/arXiv.2301.09105). The argumentation in that appendix mixes a mechanical approach where STOVs are seen in space evolving in time with an optical approach where STOVs are seen in space-time evolving with propagation distance, which leads to wrong conclusions. We consistently carry out the analysis from the mechanical approach, and the results confirm the theory in Prog. Electromagn. Res. 177, 95 consistently performed within the optical approach.
\end{abstract}

\maketitle

\begin{figure}
\includegraphics*[width=6.5cm]{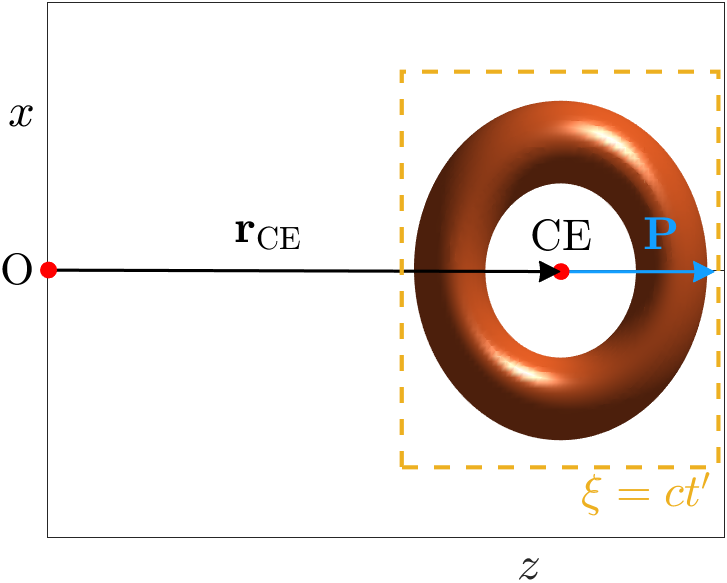}
\caption{\label{Fig1} Collage mixing properties of STOVs in $(x,y,z)$ and in $(x,y,t')$. The $y$-axis is outward from the screen.}
\end{figure}

In Ref. \cite{MILCHBERG_PRX}, page 11, a scheme similar to that in Fig. \ref{Fig1} is used to infer that the intrinsic transverse orbital angular momentum (OAM) about a $y$ axis (perpendicular to the screen) passing through the center of the energy ${\bf r}_{\rm CE}$ of a symmetric STOV propagating along the $z$ direction is equal to the transverse OAM about a $y$ axis passing though the origin O. Indeed, the extrinsic transverse OAM, ${\bf r}_{\rm CE}\times {\bf P}$, seems to vanish. This is in contradiction with Ref. \cite{PORRAS_PIERS} demonstrating that in this same problem the transverse OAM about O vanishes, and the extrinsic and intrinsic transverse OAM do not, but are opposite. Expressed per unit energy $W$, Ref. \cite{PORRAS_PIERS} demonstrates that
\begin{equation}\label{RIGHT}
   \frac{J_y}{W} = 0 \,, \quad  \frac{J_y^{(e)}}{W}=-\frac{l}{2\omega_0}\gamma\,, \quad \frac{J_y^{(i)}}{W} =\frac{l}{2\omega_0}\gamma\,,
\end{equation}
where $\omega_0$ is the carrier frequency, $l$ is the topological charge of the STOV at the transversal plane where it is elliptical with ellipticity $\gamma$. 
 
The fact that the transverse OAM is different about the moving $y$ through CE and about the fixed $y$ axis through O is not evident. Angular momentum issues are often very subtle. Understanding why they are different will inform us of the actual nature of the OAM of STOVs ---a hot topic in the structured-light community already moving towards applications \cite{HUANG}, and incidentally will reveal details of the spatial structure of STOVs that have gone unnoticed.

The error in the analysis in Ref. \cite{MILCHBERG_PRX} and in Fig. \ref{Fig1} comes from confusing 1) the structure of the STOV in space $(x,y,z)$ at each instant of time, as usually viewed with moving objects in mechanics, i.e., a snapshot, with 2) the structure of the STOV in space-time $(x,y,t')$, where $t'=t-z/c$ is the local time, at each transversal plane, as usually observed in optics (in Refs. \cite{MILCHBERG_PRX, MILCHBERG_PRL}, $\xi=ct- z =ct'$ proportional to $t'$ is used to endow the local time with dimensions of length). This confusion leads to the ``collage" of Fig. \ref{Fig1}, or of Fig. 7(b) in Ref. \cite{MILCHBERG_PRX}, where the structure in $(x,y,z)$ at an instant of time is mixed with the structure in $(x,y,t')$ at a transversal section $z$. In Ref. \cite{PORRAS_PIERS} the analysis is consistently performed in space-time $(x,y,t')$ at each $z$.

Specifically, the intrinsic OAM, or OAM about the center of the energy, is written in \cite{MILCHBERG_PRX} as
\begin{eqnarray}
{\bf J}^{(i)} &=& \int d^3{\bf r} ({\bf r}-{\bf r}_{\rm CE})\times {\bf p}\nonumber\\
              &=&  \int d^3{\bf r} \,{\bf r}\times {\bf p} -{\bf r}_{\rm CE}\times \int d^3{\bf r} \,{\bf p} \nonumber \\
              &=& \int d^3{\bf r} \,{\bf r}\times {\bf p} -{\bf r}_{\rm CE}\times {\bf P} \label{JI}
\end{eqnarray}
where $d^3{\bf r}=dxdydz$, ${\bf r}_{\rm CE}$ is the center of the energy of the STOV, ${\bf p}={\bf S}/c^2$ is the momentum density, and ${\bf S}$ is the Poynting vector ${\bf S}=\mu^{-1}_0 {\bf E}\times {\bf B}$. The first term in Eq. (\ref{JI}) is the angular momentum about the origin O, and therefore the second term is the extrinsic angular momentum of the STOV about O. Then, inspection of Fig. \ref{Fig1} ``evidences" that ${\bf r}_{\rm CE}= c(t-t_0)\hat{\bf z}$, where $\hat{\bf z}$ is the unit vector along the $z$ direction, and the total momentum ${\bf P}$ are parallel, concluding that the extrinsic OAM ${\bf r}_{\rm CE}\times {\bf P}$ vanishes.

\begin{figure*}
\includegraphics*[width=4cm]{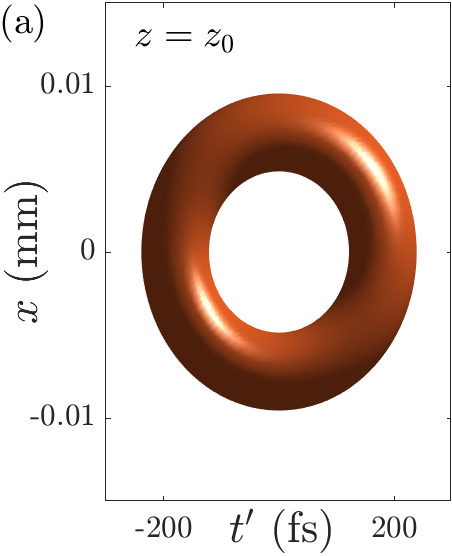}\hspace{5mm}\includegraphics*[width=6.5cm]{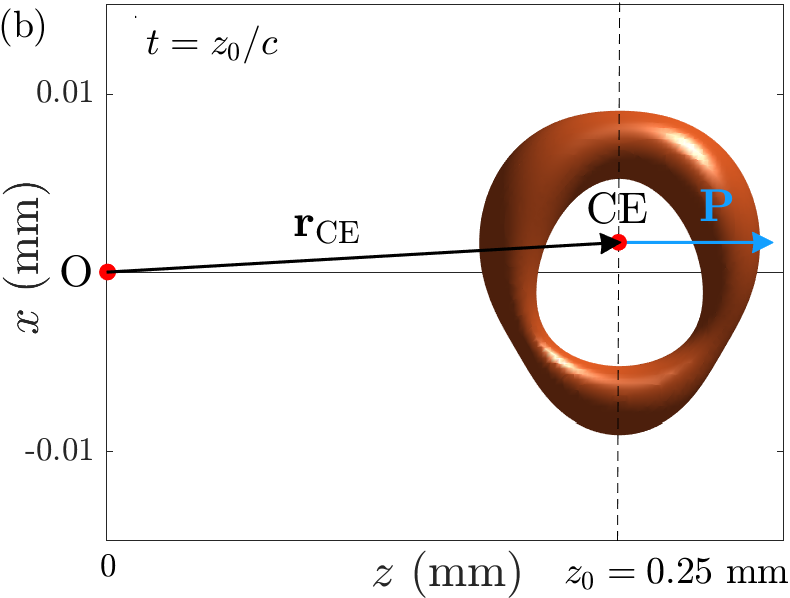}
\caption{\label{Fig2} Iso-energy-density surfaces ($80\%$ of the maximum value) (a) in $(x,y,t')$ at $z=z_0$ and (b) in $(x,y,z)$ at $t=z_0/c$ of the STOV in Eq. (\ref{STOVZ}) with $\omega_0=2.8$ rad/fs, $l=1$, $w_0=0.01$ mm, $t_0=250$ fs (ellipticity $\gamma=ct_0/w_0=7.5$). The $y$-axis is outward from the screen.}
\end{figure*}

However, when properly plotted without mixing the $(x,y,t')$ and the $(x,y,x)$ structures, ${\bf r}_{\rm CE}$ has an $x$ component. Figure \ref{Fig2}(a) shows a surface of constant energy density of an elliptical STOV in $(x,y,t')$ at a certain plane $z_0$ (see caption for the parameters defining the STOV). For consistency with the mechanical view, we look now at Fig. \ref{Fig2}(b), where a surface of the same constant energy density of the same STOV in $(x,y,z)$ is plotted at the instant of time of arrival $t=z_0/c$ at the plane $z_0$, and where it is clear (as also calculated below) that the center of the energy ${\bf r}_{\rm CE}$ has an $x$ component. Since ${\bf r}_{\rm CE}$ and ${\bf P}$ are not parallel, the extrinsic OAM ${\bf r}_{\rm CE}\times {\bf P}$ of the STOV about the $y$ axis through the origin O does not vanish. Of course the same happens with $z_0=0$, but we have chosen a positive value to imitate Fig. 7(b) of Ref. \cite{MILCHBERG_PRX}. Also, we have chosen parameters that make the non-ellipticity evident, but as a matter of fact, perfect elliptical STOVs in space $(x,y,z)$ do not exist. For positive $l$, $x_{\rm CE}$ is positive, yielding a negative extrinsic transverse OAM with respect to O. For negative $l$, the energy pattern is inverted vertically and $x_{\rm CE}$ is negative, yielding a positive extrinsic transverse OAM. This all fits with Eqs. (\ref{RIGHT}) from Ref. \cite{PORRAS_PIERS}.

The interested readers can follow the Supplementary Material below detailing how the energy density surface in Fig. \ref{Fig2}(b), the quantitative values of the center of the energy and of the total, extrinsic and intrinsic transverse OAM are calculated consistently with the mechanical approach by always integrating the densities in full space. For example, for the STOV of Fig. \ref{Fig2} at the instant of time $t=z_0/c$ of arrival at $z_0$, the energy center is calculated to be $x_{\rm CE} = 0.40018\times 10^{-3}$ mm, $y_{\rm CE}=0$ mm, and $z_{\rm CE}=ct =z_0=0.55$ mm, the OAM about O is found to be $J_y/W= 0$ fs, the extrinsic OAM is $J_y^{(e)}/W=-1.339$ fs, and hence the intrinsic OAM is $J_y^{(i)}/W=1.339$ fs. At other times, $x_{\rm CE}$ and $y_{\rm CM}$ are the same, $z_{\rm CE}=ct$, and $J_y$, $J_y^{(e)}$ and $J_y^{(i)}$ are verified to be conserved. The above values of the total, extrinsic  and intrinsic OAM coincide with those provided by Eqs. (\ref{RIGHT}) from Ref. \cite{PORRAS_PIERS} since $(l/2\omega_0)\gamma= 1.339$ fs for this STOV.

Work partially supported by the Spanish Ministry of Science and Innovation under Contract No. PID2021-122711NB-C21.

\section*{Supplementary material}

A quantitative evaluation of the energy density pattern, the surface in Fig. \ref{Fig2}(b), the center of the energy and the transverse OAM in a consistent way with the mechanical picture requires the knowledge of the full $(x,y,z)$ structure of the STOV. We start with a STOV of scalar field $\psi(x,y,t')e^{-i\omega_0 t'}$,
\begin{equation}\label{STOV0}
  \psi(x,y,t') = e^{-\frac{y^2}{w_0^2}}e^{-\frac{x^2}{w_0^2}} e^{-\frac{t^{\prime 2}}{t_0^{2}}}\left(\frac{t^{\prime}}{t_0}\mp i\mbox{sign}(l) \frac{x}{w_0}\right)^{|l|} \,,
\end{equation}
at a plane $z_0$, whose energy density is elliptical in $(x,y,t')$ with ellipticity $\gamma\equiv ct_0/w_0$, as in Fig. \ref{Fig2}(a). Its forward and backward propagated field $\psi(x,y,t',z)e^{-i\omega_0 t'}$ has been evaluated in \cite{PORRAS_OL} (and in \cite{MILCHBERG_PRL} for $|l|=1$). Once $t'$ is replaced with $t-z/c$, the STOV field reads as $\psi(x,y,z,t)e^{-i[\omega_0(t-z/c)]}$, where
\begin{equation}\label{STOVZ}
\begin{split}
  &\psi(x,y,z,t) = \frac{-iz_R}{q(z)} e^{\frac{ik_0y^2}{2q(z)}} e^{\frac{ik_0x^2}{2q(z)}}e^{-\frac{(t-z/c)^2}{t_0^2}} \left(\frac{z-z_0}{q(z)}\right)^{\frac{|l|}{2}} \\
  &\times   \frac{1}{2^{|l|}}H_{|l|}\!\left\{\!\left(\frac{q(z)}{z-z_0}\right)^{\frac{1}{2}}\!\!\left[\frac{t-z/c}{t_0} \mp \mbox{sign}(l) \frac{x}{w_0}\frac{z_R}{q(z)} \right]\!\right\}
\end{split}
\end{equation}
$k_0=\omega_0/c$, $q(z)=(z-z_0)-iz_R$, $z_R=k_0w_0^2/2$, and $H_{|l|}(\cdot)$ is the Hermite polynomial of order $|l|$. From Eq. (\ref{STOVZ}), one can construct paraxial and quasi-monochromatic electromagnetic fields following Lax's {\it et al.} perturbation theory \cite{LAX} as
\begin{eqnarray}\label{FIELDS}
\begin{split}
E_y =& {\rm Re}\left\{\psi e^{-i\omega_0 \left(t-\frac{z}{c}\right)}\right\}, \\ E_z=&{\rm Re}\left\{\frac{i}{k_0}\partial_y \psi e^{-i\omega_0 \left(t-\frac{z}{c}\right)}\right\}, \\
B_x=&-{\rm Re}\left\{\frac{1}{c}\psi e^{-i\omega_0 \left(t-\frac{z}{c}\right)}\right\}, \\ 
B_z=&-{\rm Re}\left\{\frac{i}{k_0c}\partial_x \psi e^{-i\omega_0 \left(t-\frac{z}{c}\right)}\right\}.
\end{split}
\end{eqnarray}
Linear polarization along the $y$ direction is chosen, but the results of the following calculations are the same for $x$ polarization. 

The electromagnetic energy density $w=(1/2)\left(\varepsilon_0|{\bf E}|^2 + \mu_0^{-1}|{\bf B}|^2\right)$ of the fields in Eqs. (\ref{FIELDS}) averaged over the optical cycles yields $\langle w\rangle= \varepsilon_0|\psi|^2/2$.
Note that $\langle w\rangle$ and the intensity $\langle S_z\rangle =\epsilon_0 c|\psi|^2/2$ are proportional in the paraxial approximation so the choice of one or another, or simply $|\psi|^2$, is irrelevant to determine their distributions in space. The center of energy is
\begin{equation}\label{CE}
{\bf r}_{\rm CE}(t)=\frac{1}{W}\int d^3{\bf r} \,{\bf r} \langle w\rangle = \frac{\int d^3{\bf r}  \,{\bf r} |\psi|^2}{\int d^3{\bf r}|\psi|^2} \,,
\end{equation}
where
\begin{equation}\label{E}
W=\int d^3{\bf r} \, \langle w \rangle =\frac{\varepsilon_0}{2}\int d^3{\bf r} \, |\psi|^2
\end{equation}
is the energy carried by the STOV. 

Let us continue with the $y$ component of the OAM. Its density $j_y =(zp_x-xp_z) = (zS_x-xS_z)/c^2 = \varepsilon_0 (z E_y B_z + x E_y B_x)$ time-averaged over the optical cycles and integrated in space yields the transverse OAM about O as
\begin{equation}\label{OAM}
  J_y= \frac{\varepsilon_0}{2\omega_0}\int d^3{\bf r} \left(z {\rm Im} \left\{\psi^{\star} \partial_x\psi\right\} -k_0 x |\psi|^2 \right).
\end{equation}
Similarly, the extrinsic transverse OAM $J_y^{(e)}=[{\bf r}_{\rm CE}(t)\times {\bf P}]_y = z_{\rm CE}(t) P_x - x_{\rm CE}(t) P_z$ yields
\begin{equation}\label{OAME}
J_y^{(e)} =z_{\rm CE}(t) \frac{\varepsilon_0}{2\omega_0} \int d^3{\bf r} {\rm Im}\left\{\psi^{\star}\partial_x\psi\right\} - x_{\rm CE}(t)\frac{\varepsilon_0}{2c}\int d^3{\bf r} |\psi|^2 ,
\end{equation}
and the intrinsic transverse OAM is calculated as $J_y^{(i)}=J_y-J_y^{(e)}$.

None of these integrals in $x,y,z$ can be performed analytically, given the complexity of the expression of the STOV in full space in Eq. (\ref{STOVZ}), but the numerical results corroborate Eqs. (\ref{RIGHT}), as pointed out above for the STOV of Fig. \ref{Fig2}.

\end{document}